\documentclass{aa}  

\usepackage{graphicx}
\usepackage{txfonts}
\usepackage{amsmath,amssymb}
\begin{document}

   \title{Significance testing for quasi-periodic pulsations in solar and stellar flares}


   \author{C. E. Pugh\inst{1}
          \and
          A.-M. Broomhall\inst{1,2}
          \and
          V. M. Nakariakov\inst{1,3}
          }

   \institute{Department of Physics, University of Warwick,
              Coventry, CV4 7AL, UK\\
              \email{c.e.pugh@warwick.ac.uk}
         \and
             Institute of Advanced Study, University of Warwick, 
             Coventry, CV4 7HS, UK
         \and
             St. Petersburg Branch, Special Astrophysical Observatory, 
             Russian Academy of Sciences, 196140, St. Petersburg, Russia
             }

   \date{Received February 10, 2017; accepted March 20, 2017}

 
  \abstract{The robust detection of quasi-periodic pulsations (QPPs) in solar and stellar flares has been the topic of recent debate. In light of this, we have adapted a method described by \citet{2005A&A...431..391V} to aid with the search for QPPs in flare time series data. The method identifies statistically significant periodic signals in power spectra, and properly accounts for red noise as well as the uncertainties associated with the data. We show how the method can be further developed to be used with rebinned power spectra, allowing us to detect QPPs whose signal is spread over more than one frequency bin. An advantage of these methods is that there is no need to detrend the data prior to creating the power spectrum. Examples are given where the methods have been applied to synthetic data, as well as real flare time series data with candidate QPPs from the Nobeyama Radioheliograph. These show that, despite the transient nature of QPPs, peaks corresponding to the QPPs can be seen at a significant level in the power spectrum without any form of detrending or other processing of the original time series data, providing the background trends are not too steep.}


   \maketitle
%

\section{Introduction}

Quasi-periodic pulsations (QPPs) have been widely observed in solar flares \citep[e.g.][]{2015SoPh..290.3625S, 2016ApJ...827L..30H, 2016SoPh..291.3439M} after they were first discovered by \citet{1969ApJ...155L.117P}, and they are also occasionally observed in stellar flares \citep[e.g.][]{2003A&A...403.1101M, 2016MNRAS.459.3659P}. Since QPPs are a common feature of flares, the nature of them should be understood in order to fully understand flares. It is thought that QPPs could be the result of magnetohydrodynamic oscillations or a regime of repetitive magnetic reconnection: also referred to as `load/unload' mechanisms or `magnetic dripping', where free magnetic energy continuously builds up but is released repetitively each time some threshold energy is surpassed \citep{2009SSRv..149..119N, 2016SSRv..200...75N, 2016SoPh..291.3143V}. The `magnetic dripping' term arises because an analogy can be made between this mechanism and leaking water accumulating at a steady rate at the bottom of a surface, which drips each time the weight of the water is great enough to overcome the surface tension. Although there is no strict definition of QPPs, it is generally accepted for stationary QPPs that the impulsive and/or decay phase of the flare should contain, at the very least, three cycles of oscillation, or pulses with approximately equal time spacing, visible above the noise level. There may also be non-stationary QPPs, where the time spacing between pulses increases or decreases in a non-random fashion \citep[e.g.][]{2010SoPh..267..329K, 2014ApJ...791...44H}.

The most commonly used methods to test for the presence of QPPs in a flare involve looking for a significant peak in the periodogram or wavelet spectrum. Early examples of these methods are shown by \citet{1978SoPh...57..191L, 1998ApJ...505..941A}, and for more recent examples see \citet{2011A&A...525A.112R, 2015AdSpR..56.2769C, 2016ApJ...822....7K, 2016ApJ...823L..16T}. The periodogram or wavelet spectrum of the unmodified flare light curve will, however, have power that is dependent on the frequency, following a power law relationship ($P \propto f^{-\alpha}$). This spectral behaviour is the result of red noise in the light curve, which is intrinsic to flare time series data \citep{2007ApJ...662..691M, 2011A&A...533A..61G}. In order to remove this red noise behaviour and make peaks due to a periodic component of the signal more prominent in the power spectrum, the flare light curve is often detrended. This is equivalent to the separation of the different physical phenomena of different time scales, and hence seems to be well justified. By removing the longer time-scale flare profile from the time series data, only the shorter time-scale QPPs (if present) and noise should be left. This is usually done by subtracting either a model fitted to the flare profile \citep[e.g.][]{2013ApJ...773..156A}, a boxcar smoothed version of the time series with a pre-selected boxcar width \citep[e.g.][]{2013SoPh..284..559K}, or an aperiodic trend determined by empirical mode decomposition \citep[e.g.][]{2016ApJ...830..110C}. It has been noted, however, that detrending can lead to overestimating the significance of peaks in the power spectrum \citep{2015ApJ...798..108I} due to the artificial suppression of other spectral components \citep{2011A&A...533A..61G}. A more serious potential consequence of detrending with an inappropriate model or a boxcar smooth is the introduction of a signal that looks periodic, but does not exist in the original data \citep{2016ApJ...825..110A}. 

In terms of assessing the significance of a peak in a power spectrum, \citet{1982ApJ...263..835S} showed how the false alarm probability can be found for data with white noise. For evenly time-spaced data, the periodogram is equivalent to the Fourier power spectrum (with additional normalisation), which is equal to the sum of the squares of the real and imaginary parts of the Fourier transform. For a white noise time series, where each value is drawn at random from a Gaussian distribution, the real and imaginary parts of the Fourier transform should also be Gaussian distributed random variables. Squaring a Gaussian distributed random variable results in a chi-squared, one degree of freedom (d.o.f.) distributed random variable, and adding together two chi-squared, one d.o.f. distributed variables results in a chi-squared, two d.o.f. distributed random variable. The probability density of a chi-squared, two d.o.f. distributed variable, $x$, which has a mean value of two is:
	\begin{equation}
		p_{\chi ^2_2}(x) = \frac{1}{2}\mathrm{e}^{-x/2}\,.
	\end{equation}
The probability of having a value $X$ that is greater than some threshold $x'$ is therefore:
	\begin{equation}
		\text{Pr}\left\{X>x'\right\} = \int_{x'}^{\infty} \frac{1}{2}\mathrm{e}^{-x/2} \mathrm{d}x = e^{-x'/2}\,.
	\end{equation}
Since the power spectrum is positive everywhere, $x$ and $x'$ in the above equation are always positive. When considering a power spectrum sampled at $N' = N - 1$ independent frequencies \citep[the Nyquist frequency is neglected as it follows a chi-squared one d.o.f. distribution;][]{2005A&A...431..391V} the probability is equivalent to:
	\begin{equation}
		\text{Pr}\left\{X>x'\right\} = 1 - \left(1 - \epsilon _{N'}\right)^{1/N'} \approx \epsilon _{N'}/N'\,,
	\end{equation}
where $\epsilon _{N'}$ is the false alarm probability, and the approximation holds when is $\epsilon _{N'}$ small \citep{2002MNRAS.336..979C}. The false alarm probability is defined as the probability of observing a peak in the power spectrum above some threshold power \citep{1982ApJ...263..835S, 1986ApJ...302..757H}, and this threshold power can be calculated by equating Eqs. (2) and (3) and solving for $x'$. For example, the 99\% confidence level in the power spectrum is the power threshold corresponding to a false alarm probability of 0.01, and refers to the level above which there is only a 1\% chance of observing a peak in the power spectrum of a Gaussian distributed random (white noise) time series. The above expressions are only valid, however, when considering data with white noise that is chi-squared, two d.o.f. distributed in the power spectrum. When considering solar and stellar flare time series data, the removal of an imperfect approximation of the underlying flare trend could alter the underlying noise distribution, leading to the calculation of a misleading confidence level. Therefore extreme care should be taken when assessing the significance of peaks in the power spectrum of detrended data.

Another point to consider when detrending, is that each flare must be treated separately in the analysis. For example, a model that gives a good fit to the flare profile must be chosen, along with suitable initial estimate parameters, but some flares are very complex in shape \citep[e.g.][]{2014ApJ...797..122D}, which makes finding a general model that fits all flares to a satisfactory standard more difficult. In addition to this, the same flare can look very different when viewed in different energy bands. On the other hand, if a boxcar smooth is used to detrend the most suitable boxcar width must be chosen for each flare. When undertaking a large-scale study of a number of events a minimal amount of manual intervention is preferable, hence methods that avoid detrending are more appropriate, such as that used by \citet{2016ApJ...833..284I}. 

\citet{2005A&A...431..391V} demonstrates a method to assess the significance of a peak in a power spectrum, and shows how it can be used to test for periodic signals in X-ray light curves of active galaxies. The method avoids detrending and takes full account of red noise and data uncertainties. We build on this method and show in detail how it can be applied to solar flare data, so that peaks in the power spectrum found above a certain confidence level may be considered as candidate QPPs. For this study we address only stationary QPPs, with constant periods and without phase modulation. In Section 2 we summarise the method of \citet{2005A&A...431..391V}, derive a more simple form of the equation to be solved to determine the confidence level, and describe how this method can be used to search for QPPs in flare light curve data. In Section 3 we adapt the method to be used with rebinned power spectra, which can help detect QPPs with a broad peak in the power spectrum. The results of testing the methods on simulated data are shown in Section 4, and in Section 5 a few examples are given where the methods have been applied to real solar flare data. A summary is given in Section 6.


\section{Confidence levels on power-law power spectra}

From \citet{2005A&A...431..391V}, if $I(f_j)$ is the periodogram power at a particular frequency, $f_j$, we have:
	\begin{equation}
		I(f_j) = \mathcal{P}(f_j)\chi ^2_2/2\,,
	\end{equation}
where $\mathcal{P}(f_j)$ is the `true' power, and $\chi ^2_2/2$ is the chi-squared two d.o.f. distributed noise. For data with red noise, the power spectrum will follow a power law, which can be fitted with a straight line when working in log space. Solar flare power spectra often follow a broken power law, since the red noise component can fall below the white noise level at high frequencies \citep{2011A&A...533A..61G, 2016SSRv..198..217M}. The broken power law model can be written as:
\begin{equation}
	\log\left[\hat{\mathcal{P}}(f)\right] = 
		\begin{cases}
			-\alpha\log\left[f\right] + c & \text{if } f < f_{break} \\
			-\left(\alpha - \beta\right)\log\left[f_{break}\right] - \beta\log\left[f\right] + c & \text{if } f > f_{break}\,,
		\end{cases}
\end{equation}
where $f_{break}$ is the frequency at which the power law break occurs, $\alpha$ and $\beta$ are power law indices and $c$ is a constant. The probability density for ${2I_j}$ (the factor of two appears because the chi-squared distribution is conventionally defined assuming the values following that distribution have a mean equal to the number of degrees of freedom) can then be written as:
	\begin{equation}
		p_{2I_j}(x) = \frac{1}{2\mathcal{P}_j}\mathrm{e}^{-x/2\mathcal{P}_j}\,.
	\end{equation}

When fitting the `true' power spectrum, $\mathcal{P}$, with a broken power law model, $\hat{\mathcal{P}}$, the uncertainties on this fitted model will follow a Gaussian distribution in log space, and hence a log-normal distribution in linear space:
	\begin{equation}
		p_{\hat{\mathcal{P}}_j}(y) = \frac{1}{\sqrt{2\pi\;}yS_j}\exp{\left\{-\frac{\left(\ln[y] - \ln[\mathcal{P}_j]\right)^2}{2S^2_j}\right\}}\,,
	\end{equation}
with
	\begin{equation}
		S_j = \text{err}\left\{\log\left[\hat{\mathcal{P}}(f_j)\right]\right\} \times \ln[10]\,,
	\end{equation}
where $\text{err}\left\{\log\left[\hat{\mathcal{P}}(f_j)\right]\right\}$ is the uncertainty on the fitted model in log space, and the $\ln[10]$ factor accounts for the fact that the uncertainty is defined in terms of log base ten, whereas the log-normal distribution above is defined (by convention) in terms of log base $e$. In order to find the uncertainties on the model fitted to flare power spectra, uncertainties on the flare light curve data were used in Monte Carlo simulations. While some instruments that observe the Sun include uncertainties for the data provided, many do not. Fortunately in most cases reasonable estimates of the uncertainties can be made. For example, some X-ray observations of the Sun follow Poisson counting statistics, so the uncertainty on each count rate value can be found by taking the square root of the value. For the radioheliograph data used in Section 5, an estimate of the uncertainty can be found by calculating the standard deviation of several hours of flat data, in which no flares occur and there are no other features. For the Monte Carlo simulations, random numbers with a mean of zero and standard deviations equal to the uncertainties on the light curve values are added to each of the light curve values. The periodogram is then found, converted to log space, and a broken power law model is fitted using a least-squares method. This is repeated many times (10,000 times for the examples shown in Section 5), and for each iteration the initial guess parameters used in the model fit are allowed to vary, in order to prevent a local rather than a global minimum being found by the least-squares fit (for the examples in Section 5 parameters of $\alpha = 2$, $\beta = 1$, $c = -3$ and $f_{break} = 0.1$ were allowed to randomly vary with a standard deviation equal to 10\% of the parameter value for each iteration). The distributions of the parameters from the repeated power spectrum fits should be approximately Gaussian, and so for each frequency bin of the power spectrum, the uncertainty on the fitted model value will be Gaussian distributed. Hence the distribution of fitted powers at each frequency index can be fitted by a Gaussian model, and the standard deviation of the Gaussian model can be used as an estimate of the uncertainty of the broken power law model at that index.

\begin{figure}
        \centering
        \includegraphics[width=\linewidth]{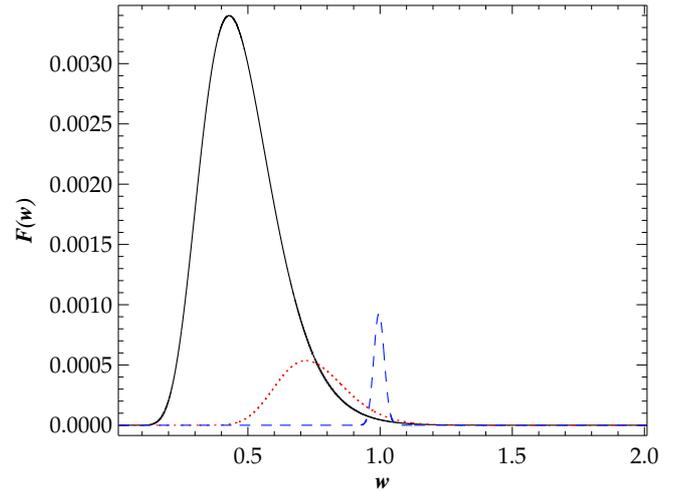}
        \caption{Plots of the integrand in Eq. (12) as a function of $w$. The solid black, dotted red and dashed blue lines show the function when $S_j$ is equal to 0.4, 0.2 and 0.02 respectively. The value of $\gamma _{\epsilon _j}$ has arbitrarily been chosen to be equal to 20, which is a typical value.}
        \label{int_func}
   \end{figure}

\begin{figure}
        \centering
        \includegraphics[width=\linewidth]{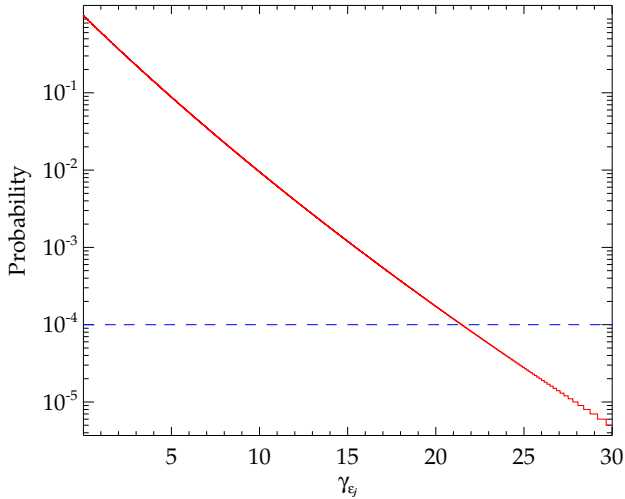}
        \caption{Plots of Eq. (12) and Eq. (13) as a function of $\gamma _{\epsilon _j}$ are shown by the solid red and dashed blue lines respectively. The values of $S_j$ and $\epsilon _{N'}/N'$ have arbitrarily been chosen to be equal to 0.2 and 0.01/100 respectively. The solution we require is where Eq. (12) is equal to Eq. (13), which corresponds to $\gamma _{\epsilon _j} = 21.467$.}
        \label{int_sol}
   \end{figure}

The probability density of the ratio $\hat{\gamma}_j = 2I_j/\hat{\mathcal{P}}_j$ (the power spectrum with the red noise component removed) can be found from \citep{curtiss1941}:
	\begin{equation}
		p_{\gamma_j}(z) = \int_{0}^{+\infty} |y|p_{2I_j}(zy)p_{\mathcal{P}_j}(y) \mathrm{d}y\,,
	\end{equation}
where $y$ and $z$ are dummy variables representing different power levels in the power spectrum. The lower limit of this integral is zero rather than negative infinity because the power spectrum is always positive. Integrating this probability density between $\gamma _{\epsilon _j}$ and infinity gives the probability that a value $\hat{\gamma _j}$ is greater than $\gamma _{\epsilon _j}$:
	\begin{equation}
	\resizebox{.49\textwidth}{!}
	{ 
		$ \text{Pr}\left\{\hat{\gamma _j} > \gamma _{\epsilon _j}\right\} = \frac{1}{\sqrt{8\pi\;}S_j\mathcal{P}_j}\int_{\gamma _{\epsilon _j}}^{\infty}\int_{0}^{\infty}\exp \left\{-\frac{\left(\ln[y] - \ln[\mathcal{P}_j]\right)^2}{2S^2_j} - \frac{yz}{2\mathcal{P}_j}\right\} \mathrm{d}y\mathrm{d}z\,.$
	}
	\end{equation}
Substituting in another dummy variable, $w = y/\mathcal{P}_j$ (with $\mathrm{d}y = \mathcal{P}_j\;\mathrm{d}w$), simplifies this to:
	\begin{equation}
		\text{Pr}\left\{\hat{\gamma _j} > \gamma _{\epsilon _j}\right\} = \frac{1}{\sqrt{8\pi\;}S_j}\int_{\gamma _{\epsilon _j}}^{\infty}\int_{0}^{\infty}\exp \left\{-\frac{\left(\ln\,w\right)^2}{2S^2_j} - \frac{wz}{2}\right\} \mathrm{d}w\mathrm{d}z\,.
	\end{equation}
Since the integrand is well-behaved and contains no discontinuities, the order of integration can be swapped (see Fig. \ref{int_func} for plots of the function for different values of $S_j$). Then the function can be integrated with respect to $z$, to get:
	\begin{equation}
	 \text{Pr}\left\{\hat{\gamma _j} > \gamma _{\epsilon _j}\right\} = \int_{0}^{\infty}\frac{1}{\sqrt{2\pi\;}S_jw} \exp \left\{-\frac{\left(\ln\,w\right)^2}{2S^2_j} - \frac{\gamma _{\epsilon _j}w}{2}\right\} \mathrm{d}w\,,
	\end{equation}
which can be equated to (see Eq. (3) in Section 1):
	\begin{equation}
		\text{Pr}\left\{\hat{\gamma _j} > \gamma _{\epsilon _j}\right\} \approx \frac{\epsilon _{N'}}{N'}\,,
	\end{equation}
and solved numerically in order to find a $\gamma _{\epsilon _j}$ corresponding to each value of the fitted model power spectrum, $\hat{\mathcal{P}}_j$. Figure \ref{int_sol} shows plots of Eq. (12) and Eq. (13) as a function of $\gamma _{\epsilon _j}$, and the solution is where the two lines cross. In practise it is helpful to subtract Eq. (13) from Eq. (12), then the solution will be where this function is equal to zero and a root finding algorithm can be used. The final step is to ensure correct normalisation, since the conventional form of the probability density function for a chi-squared, two d.o.f. distribution (shown in Eq. (1)) assumes the values following that distribution have a mean equal to two, which is not necessarily the case. An important point to account for is that the mean calculated in log space is not the same as the log of the mean calculated in linear space (i.e. $\langle\log\mathcal{P}_j\rangle \ne \log\langle\mathcal{P}_j\rangle$). Hence the power spectrum as well as the fit must be converted into linear space so that the mean, or expectation value, of the flattened power spectrum (denoted $\langle \mathcal{P}_j - \hat{\mathcal{P}}_j\rangle$) can be found. Therefore the confidence level in log space is then equal to $\log[\hat{\mathcal{P}}_j] + \log[\gamma _{\epsilon _j}\langle \mathcal{P}_j - \hat{\mathcal{P}}_j\rangle /2]$.

\section{Confidence levels on rebinned power spectra}

\citet{2004A&A...428.1039A} shows how rebinning the power spectrum can improve the detection of short-lived solar acoustic modes, such as wave trains with a highly modulated amplitude, which have power spread across several frequency bins. \citet{1989ARA&A..27..517V} and \citet{1993MNRAS.261..612P} also describe the use of binned or smoothed power spectra in the analysis of X-ray binaries and active galaxies exhibiting quasi-periodic oscillations. A similar method can be applied to candidate QPPs, for example those with exponential or Gaussian damping typical for solar and stellar flares \citep{2016MNRAS.459.3659P, 2016ApJ...830..110C}, or those with a small variation of the period. These QPP signals may appear as a broad peak in the power spectrum, with a power spanning more than one frequency bin, and hence considering all of the power contained within the peak (by rebinning the spectrum) rather than separately considering the power in each of the frequency bins will give a better assessment of the significance of the peak.

When summing every $n$ frequency bins, the probability density follows a chi-squared $2n$ d.o.f. distribution with a mean equal to $2n$ \citep{2003A&A...412..903A}: 
	\begin{equation}
		p_{\chi ^2_{2n}}(x) = \frac{x^{n-1}\mathrm{e}^{-x/2}}{2^n\Gamma (n)}\,,
	\end{equation}
where $\Gamma$ is the gamma function. Hence the probability distribution followed by the rebinned power spectrum is:
	\begin{equation}
		p_{2nI_j}(x) = \frac{x^{n-1}\mathrm{e}^{-x/2\mathcal{P}_j}}{2^n\mathcal{P}^n_j\Gamma (n)}\,,
	\end{equation}
where $\mathcal{P}_j$ is the `true' rebinned power spectrum which can be fitted by a power law model, $\hat{\mathcal{P}}_j$. Plugging this equation into Eq. (9), along with Eq. (7) gives:
	\begin{equation}
	\resizebox{.44\textwidth}{!}
	{ 
		$ p_{\gamma_j}(z) = \int_{0}^{\infty}\frac{(yz)^{n-1}}{2^n\mathcal{P}_j^n\Gamma (n)\sqrt{2\pi\;}S_j}\exp \left\{-\frac{\left(\ln[y] - \ln[\mathcal{P}_j]\right)^2}{2S^2_j} - \frac{yz}{2\mathcal{P}_j}\right\} \mathrm{d}y\,.$ 
	}
	\end{equation}
Integrating this probability density from $\gamma _{\epsilon _j}$ up to infinity and substituting $w = y/\mathcal{P}_j$, as before, gives:
	\begin{equation}
		\text{Pr}\left\{\hat{\gamma _j} > \gamma _{\epsilon _j}\right\} = \int_{\gamma _{\epsilon _j}}^{\infty}\int_{0}^{\infty}\frac{(wz/2)^{n-1}}{\sqrt{8\pi\;}S_j\Gamma (n)}\exp \left\{-\frac{\left(\ln\,w\right)^2}{2S^2_j} - \frac{wz}{2}\right\} \mathrm{d}w\mathrm{d}z\,.
	\end{equation}
By swapping the order of integration and letting $u=wz/2$ (hence $\mathrm{d}z = 2\mathrm{d}u/w$), this equation becomes:
	\begin{equation}
	\resizebox{.5\textwidth}{!}
	{ 
		$ \text{Pr}\left\{\hat{\gamma _j} > \gamma _{\epsilon _j}\right\} = \int_{0}^{\infty}\frac{2}{\sqrt{8\pi\;}S_j\Gamma (n)w}\exp \left\{-\frac{\left(\ln\,w\right)^2}{2S^2_j}\right\} \left\{\int_{w\gamma _{\epsilon _j}/2}^{\infty} \exp (-u)u^{n-1} \mathrm{d}u\right\}\mathrm{d}w\,,$ 
	}
	\end{equation}
which, after writing the internal integral in gamma function notation, becomes:
	\begin{equation}
	 	\text{Pr}\left\{\hat{\gamma _j} > \gamma _{\epsilon _j}\right\} = \int_{0}^{\infty}\frac{1}{\sqrt{2\pi\;}S_jw} \exp \left\{-\frac{\left(\ln\,w\right)^2}{2S^2_j}\right\} \frac{\Gamma (n, w\gamma _{\epsilon _j}/2)}{\Gamma (n)}\mathrm{d}w\,,
	\end{equation}
where $\Gamma (n, w\gamma _{\epsilon _j}/2)$ is the upper incomplete gamma function. Like before, this can be solved numerically by equating to Eq. (13), and the confidence level in log space is equal to $\log[\hat{\mathcal{P}}_j] + \log[\gamma _{\epsilon _j}\langle \mathcal{P}_j - \hat{\mathcal{P}}_j\rangle /2n]$.


\section{Testing the methods on simulated data}

\begin{figure*}
        \centering
        \includegraphics[width=0.85\linewidth]{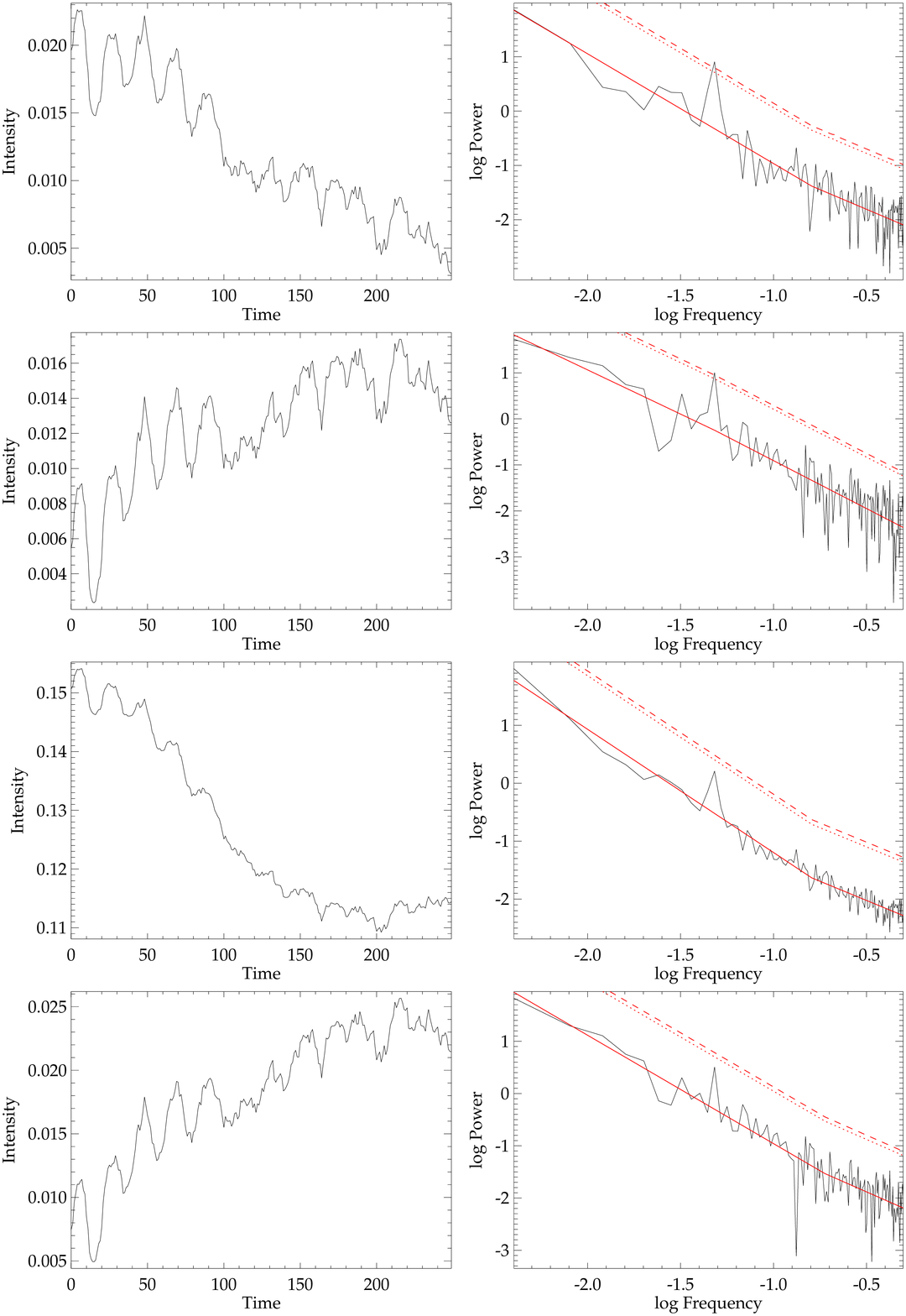}
        \caption{Examples of synthetic flare time series with QPPs are given on the left, and on the right are the corresponding power spectra, where the red solid line is a power law fit, and the red dotted and dashed lines correspond to the 95\% and 99\% confidence levels respectively. The top two rows show two signals with different background trends, both with a peak above the 99\% level in the power spectrum. The bottom two rows show the same signals but with steeper background trends, the result of which is that the peaks no longer reach significant levels in the power spectra.}
        \label{synth}
   \end{figure*}

Figure \ref{synth} shows examples where confidence levels have been calculated for synthetic flare time series with QPP signals, and shows how different background trends (which are unknown for real flare data) affect the appearance of a QPP signal in the power spectrum. To create the time series, a polynomial background trend was added to an exponentially damped sinusoid, white noise, and additional red noise. The additional red noise was generated by a random walk, where each value in the time series is equal to a random number summed with the preceding value. The parameters were chosen to be comparable to flare time series data from the Nobeyama Radioheliograph (see Section 5) and, for all of these time series, the sinusoid, white noise and random walk noise terms were kept identical. The top two rows of Fig. \ref{synth} show time series with different polynomial background trends on the left, and the corresponding power spectra on the right. Despite the different background trends, peaks corresponding to the sinusoidal signals can be seen above the 99\% confidence level in the power spectra. The bottom two rows show the same signals as the top two rows, but instead they have higher amplitude background trends. The steeper background trends mean that the sinusoidal signals are no longer seen at a significant level in the power spectra. Therefore, although the method described in Section 2 is useful for testing for the presence of a QPP signal when there is some unknown background trend in the data, when the amplitude of the background trend is sufficiently greater than the amplitude of a QPP signal, the QPP signal will be hidden in the power spectrum.

\begin{figure*}
        \centering
        \includegraphics[width=\linewidth]{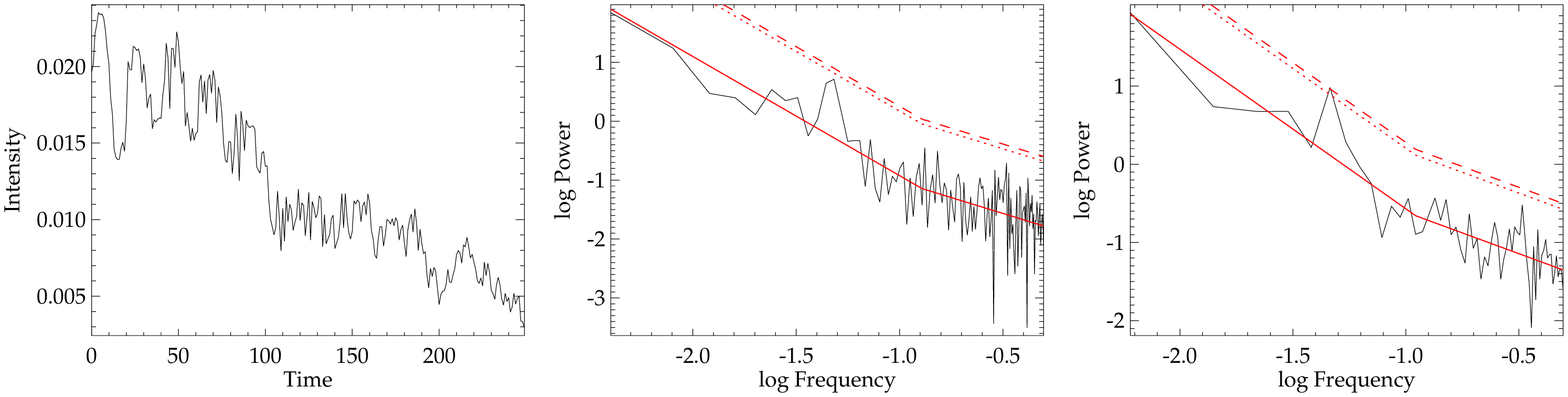}
        \caption{Example of how rebinning can help spectral peaks corresponding to certain kinds of periodic signals reach a significant power level. The left panel shows a synthetic time series signal, similar to that in the top left panel of Fig. \ref{synth}, but with a sinusoidal component that has a frequency that fluctuates slightly with time. The middle panel shows the corresponding power spectrum, and the right panel shows the rebinned power spectrum (after summing the powers in every two frequency bins). As before, the red solid line is a power law fit, and the red dotted and dashed lines correspond to the 95\% and 99\% confidence levels respectively.}
        \label{synth_rebin}
   \end{figure*}

A scenario where the method described in Section 3 results in a peak above the 99\% level in the power spectrum, while the method from Section 2 does not, is demonstrated in Fig. \ref{synth_rebin}. The left panel shows a signal with the same background and noise as the signal in the top left panel of Fig. \ref{synth}, but instead it has a sinusoidal term with a frequency that has a constant mean but a small amount of random variation with time. The result of this is that in the power spectrum, shown in the middle panel of Fig. \ref{synth_rebin}, the peak corresponding to the sinusoidal signal is spread across more than one frequency bin, and hence no longer reaches a significant power level. The rebinned power spectrum is shown in the right panel, where the powers in every two frequency bins of the original spectrum have been summed together. Here the peak corresponding to the sinusoidal signal has a power above the 99\% confidence level.

\section{Examples of application to solar flare data}

An important consideration when performing a periodogram analysis of time series data to search for a periodic signal that is localised in time (such as a wave train), is the choice of start and end times. For example, if a 5 hour section of a light curve was taken which contained a flare with QPPs that persisted for 30 minutes, then the significance of the peak in the power spectrum corresponding to the QPP signal would be lower than if a 30 minute section of the same light curve, centred around the QPP signal, was used instead. For the following solar flare light curves, the start and end times were chosen manually to best show off candidate QPPs: this was the only aspect of the analysis that was handled manually. An additional issue with the spectral analysis of time series data is the finite duration of the data. For a chosen section of a flare time series, the start and end values are unlikely to be equal to zero, hence there will be spectral leakage into side-lobes when performing some form of discrete Fourier transform, and this reduces the power of a peak in the power spectrum. One way to avoid this effect is to apply a window function, for example a Hann window, to the time series data before calculating the power spectrum. The application of such a window function is, however, not always helpful when searching for low-amplitude transient signals such as QPPs, since any QPP signals will be suppressed near the start and end of the time series, making detection even more challenging. In addition, the application of any window function other than a rectangular window will alter the distribution of noise in the data, and therefore this would need to be taken into account when using the methods described in this paper. Hence for the following examples, no window function has been applied.

\begin{figure*}
        \centering
        \includegraphics[width=\linewidth]{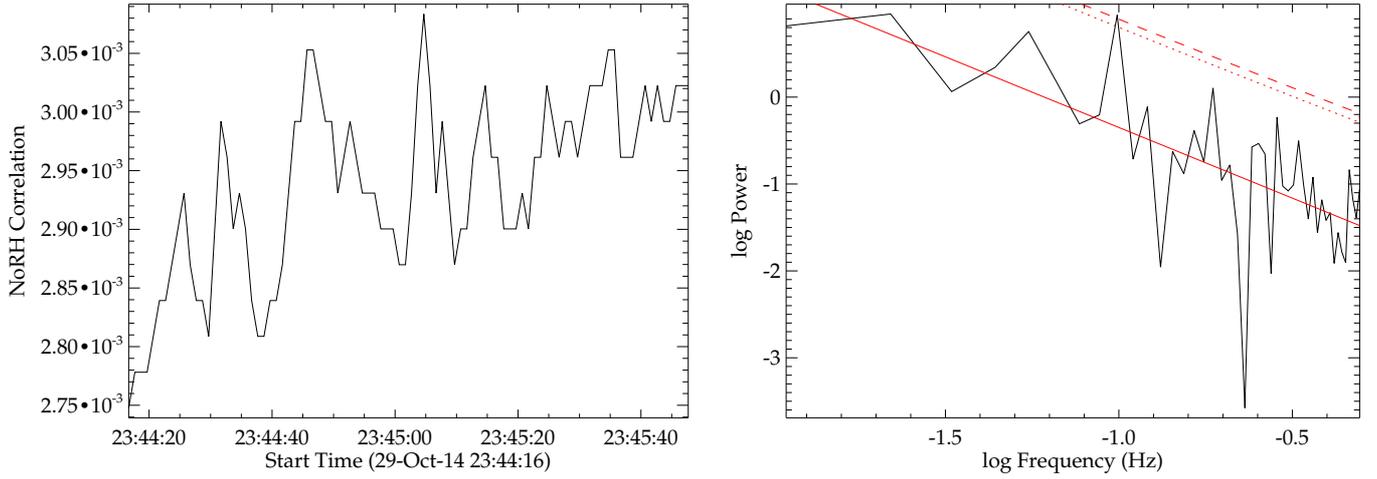}
        \caption{Left: a section of a GOES C7.1 class flare observed by Nobeyama Radioheliograph. Right: the corresponding power spectrum, where the red solid line is a power law fit to the spectrum, the red dotted line represents the 95\% confidence level, and the red dashed line the 99\% level. One peak is above the 99\% level, at a period of $10.1^{+0.6}_{-0.5}$\,s.}
        \label{138norh}
   \end{figure*}

\begin{figure*}
        \centering
        \includegraphics[width=\linewidth]{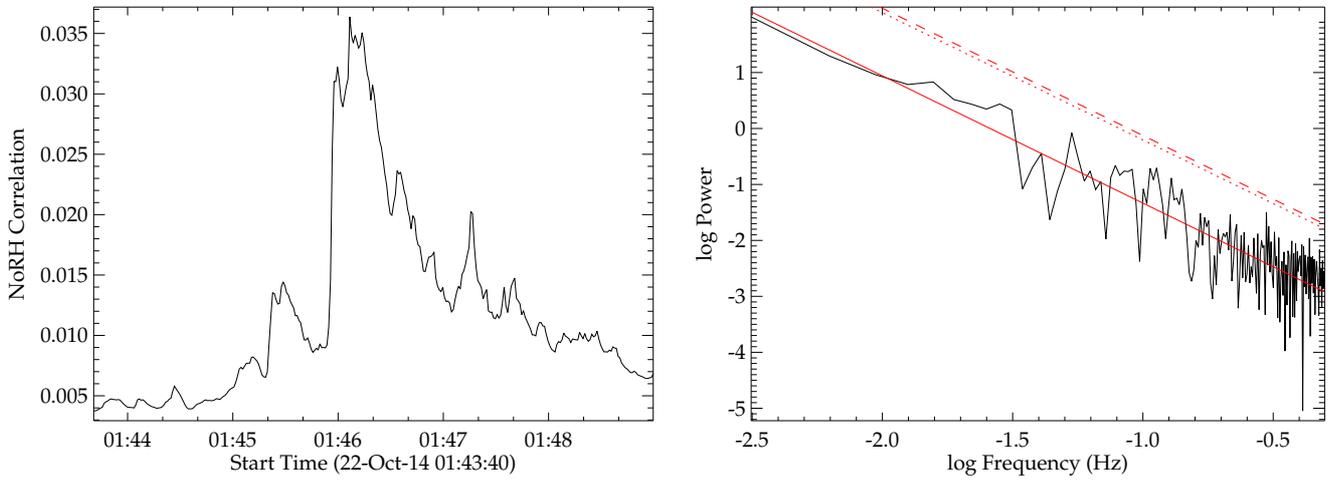}
        \caption{As in Fig. \ref{138norh}, but for a different flare with a class of M8.7. Although this flare appears to have pulsations there is no peak close to the 95\% level in the power spectrum.}
        \label{68norh}
   \end{figure*}

An example of the method described in Section 2 being used to confirm candidate QPPs in a GOES C7.1 class flare, observed between 2014 October 29 23:40 and October 30 00:34 UT, is shown in Fig. \ref{138norh}. A section of 17 GHz microwave correlation signal from the Nobeyama Radioheliograph (NoRH) \citep{1994IEEEP..82..705N} is shown in the left hand panel, and the corresponding Lomb-Scargle periodogram power spectrum is shown on the right. As mentioned in Section 2, the correlation data uncertainty was estimated by taking the standard deviation of a flat section of data, taken from 2016 October 27 00:00 until 05:00 UT. This gave an uncertainty of $1.1911749 \times 10^{-5}$. In the periodogram there is a peak with a period of $10.1^{+0.6}_{-0.5}$\,s above the 99\% confidence level, where the upper and lower uncertainties are taken to be plus or minus half a frequency bin, respectively, on either side of the peak frequency. Visual inspection of the light curve on the left confirms that the pulses have a time spacing approximately equal to this period, hence this can be considered a strong QPP candidate. An example where the method fails to support the possible presence of QPPs in a M8.7 class flare, observed between 2014 October 21 08:09 and 08:15 UT, is shown in Fig. \ref{68norh}, where there is no peak in the power spectrum above the 95\% level. Although pulsations can be seen in the light curve, these are small in amplitude when compared to the underlying trend in the data, meaning that the trend dominates in the power spectrum and even though the pulsations may be periodic they are not detectable at a significant level.

Another point to note is that although a broken power law model was used to fit these NoRH power spectra in Figs. \ref{138norh} and \ref{68norh}, the break cannot be seen. This can be explained by considering the white noise amplitude in the NoRH time series data, which is very small. Hence the frequency at which the white noise would start dominating over the red noise in the power spectrum is likely beyond the range of frequencies included in the spectrum.

\begin{figure*}
        \centering
        \includegraphics[width=\linewidth]{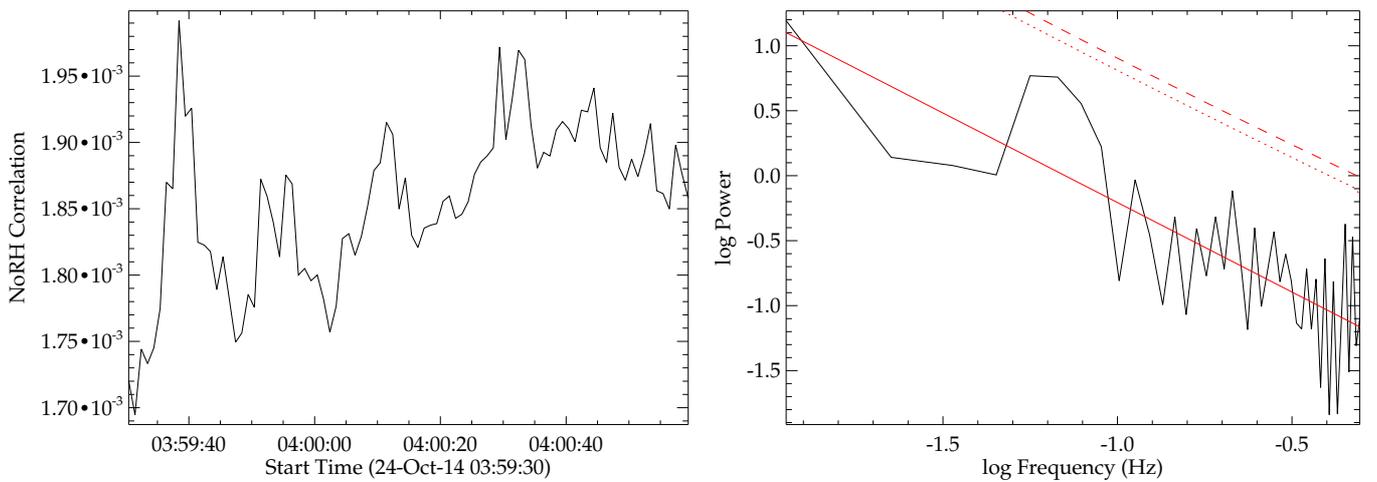}
        \caption{As in Fig. \ref{138norh}, but for a different flare with a class of C3.6. Here the power spectrum contains a broad peak, which does not reach the 95\% confidence level.}
        \label{81norh}
   \end{figure*}

\begin{figure}
        \centering
        \includegraphics[width=\linewidth]{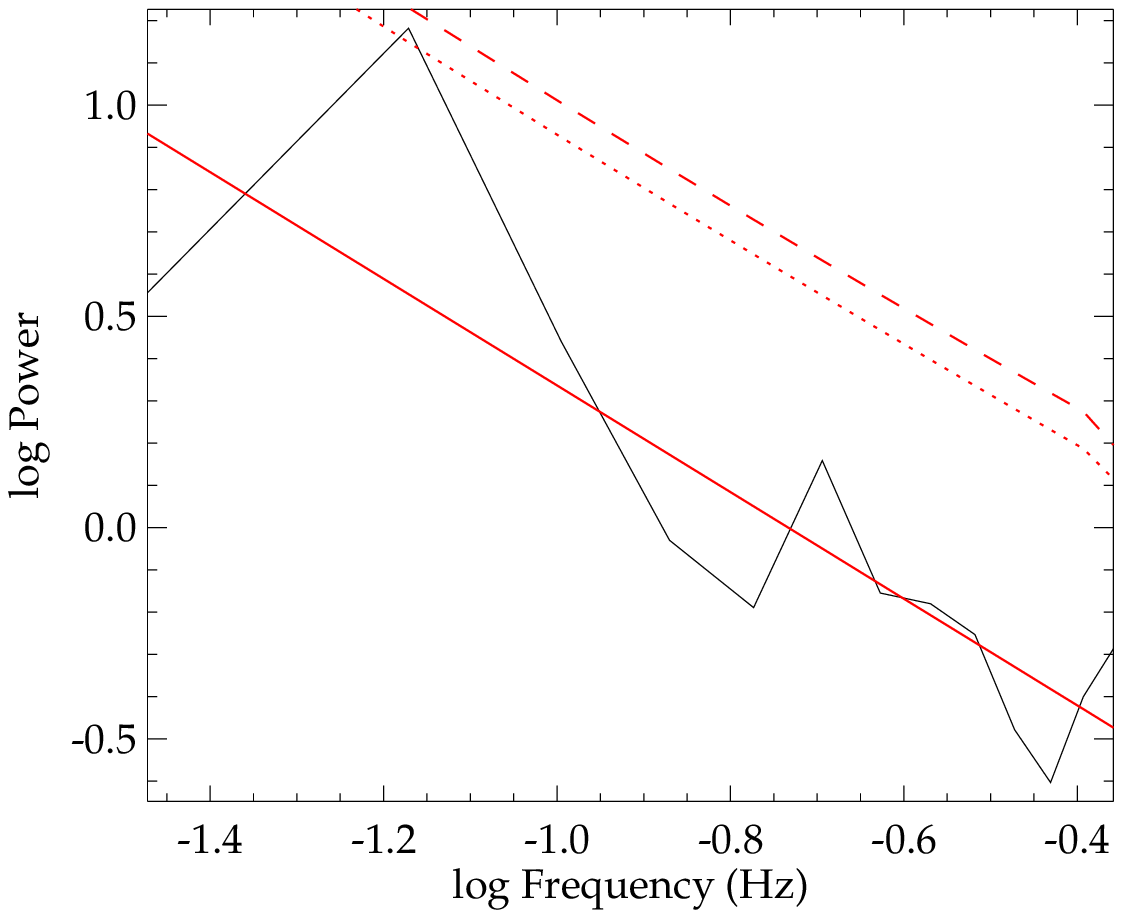}
        \caption{The rebinned power spectrum for the flare shown in Fig. \ref{81norh}. The peak at a period of $15^{+5}_{-3}$\,s now surpasses the 95\% confidence level, which is shown by the red dotted line.}
        \label{81norh_rebin}
   \end{figure}

For the section of the C3.6 class flare observed between 2014 October 24 03:56 and 04:30 UT, shown in Fig. \ref{81norh}, again the method described in Section 2 fails to show a peak in the power spectrum above the 95\% level, however a broad peak can be seen. The rebinned power spectrum is shown in Fig. \ref{81norh_rebin}, where every three frequency bins have been summed over. Applying the method described in Section 3 shows that there is a peak above the 95\% confidence level at a period of $15^{+5}_{-3}$\,s, hence this flare can be considered to have candidate QPPs.

\section{Summary}

We have demonstrated how the method of \citet{2005A&A...431..391V} can be applied to flare time series data in order to test for the presence of QPPs, subject to the careful choice of start and end times. The method has been adapted to be used with rebinned power spectra, which can aid the detection of QPP signals that have a period that varies slightly, or have a modulated amplitude. These methods avoid detrending the data, an approach which has been shown to have the potential to lead to false detections when the detrending is done by smoothing \citep[e.g.][]{2011A&A...533A..61G, 2016ApJ...825..110A}. An alternative method which also avoids detrending has been proposed by \citet{2015ApJ...798..108I}. This method involves doing a model comparison; different models, such as a power law and a power law plus Gaussian peak, are fitted to the power spectrum and are compared by calculating the Bayesian Information Criterion (BIC) for each. The model with the smaller BIC value will then be the preferred model. We therefore suggest that a thorough search for QPPs in flares could make use of both approaches. These methods, however, may not be suitable for searching for non-stationary QPP signals if the change in period is too great, in which case either some form of wavelet analysis \citep[e.g.][]{2010SoPh..267..329K} or empirical mode decomposition \citep[e.g.][]{2015A&A...574A..53K} could be used.


\begin{acknowledgements}
We would like to thank the anonymous referee for their helpful comments. C.E.P. would like to thank Alex Seaton and Mark Hollands for useful discussions. C.E.P. \& V.M.N.: This work was supported by the European Research Council under the \textit{SeismoSun} Research Project No. 321141. A.-M.B. thanks the Institute of Advanced Study, University of Warwick for their support. The authors would like to acknowledge the ISSI International Team led by A.-M.B. for many productive discussions, and are also grateful to the Nobeyama team for providing the data used.
\end{acknowledgements}


\bibliographystyle{aa}
\bibliography{ms}

\end{document}